\documentstyle[twoside,fleqn,espcrc2,epsf]{article}
\epsfverbosetrue     % report size of each figure included
\def\be{\begin{equation}}
\def\ee{\end{equation}}
\def\ba{\begin{eqnarray*}}
\def\ea{\end{eqnarray*}}

\def\L{\Lambda }

\def\A{{\cal A}}
\def\tilde#1{\widetilde{#1}}

\def\rho{\varrho}
\def\epsilon{\varepsilon}

\begin{document}
 \newcommand{\Dirac}{\not\!\nabla}
%-----------------------------------------------------------------------
%                           HEAD
%-----------------------------------------------------------------------
\title{Multigrid meets neural nets}
\author{M. B\"aker,
        G. Mack
        and M. Speh\\[3mm]
        II. Institut f\"ur Theoretische Physik,
        Universit\"at Hamburg, Luruper Chaussee 149, 2000 Hamburg 50
        }
%-----------------------------------------------------------------------
%                      Abstract
%-----------------------------------------------------------------------
\begin{abstract}
We present evidence that
multigrid (MG) works for wave equations in disordered systems, e.g. in
the presence of gauge fields, no matter how
strong the disorder. We introduce a ``neural computations''
point of view into large scale simulations: First, the system must
learn how to do the simulations efficiently, then do the simulation
(fast). The method can also
be used to provide smooth interpolation kernels which are needed in
multigrid Monte Carlo updates.
\end{abstract}
\maketitle
%-----------------------------------------------------------------------
%                      Section 1
%-----------------------------------------------------------------------
%
\section{INTRODUCTION}
   There is a stochastic multigrid method and a deterministic one.
The stochastic version is used to compute high dimensional integrals in
Euclidean quantum field theory or statistical mechanics by a Monte Carlo
method which uses updates at different length scales~\cite{1,2}. The
deterministic version~\cite{4} solves discretized partial
differential equations. One hopes to use both of them in simulations
of lattice QCD, for updating the gauge fields and for computing
fermion propagators in given gauge fields. In either case the aim is
to beat critical slowing down (CSD) in nearly critical systems.

Our notation is as follows: $\L^0 $ denotes a given
``fundamental'' lattice $\L^0$ of spacing $a_0$.
Coarser (block) lattices of increasing spacings
$a_j=L_b^ja_0$ are denoted $\L^1,\L^2,..., \L^N$.
Typically, we chose $L_b=2$, and a single point as
the last layer $\L^N$. Interpolation operators
$\A^j $ are introduced to transfer functions
on coarser lattices into functions on finer lattices,
while restriction operators $C^j \equiv \A^{j \ast }$
transfer functions from finer to coarser lattices
(``variational coarsening'').
%-----------------------------------------------------------------------
%                      Section 2
%-----------------------------------------------------------------------
%
\section{IMPORTANCE OF SMOOTHNESS \label{SMOOTHNESS}}
A crucial problem is how to define and exhibit smooth functions
in the disordered context, i.e. when translation symmetry is strongly
violated. Other possible applications besides gauge theories are
low lying states of spin glasses,
the shape of a lightning,
waves on fractal lattices (with bond percolation), or
the localization of low lying electronic states in amorphous materials.

In the case of deterministic MG,
one wants to solve a discretized elliptic differential equation on
$\L^0$:
\be D_0\xi^0 =f^0 \ .  \label{De}
\ee
It might have arisen from an eigenvalue equation
$D_0\xi^0 = \epsilon \xi^0 $ by inverse iteration.
If $D_0$ has a small
eigenvalue, then local relaxation algorithms suffer from CSD.
After some relaxation sweeps on $\L^0$ one gets an approximate
solution $\tilde{\xi }^0 $ whose error
  $ e^0 = \xi^0 - \tilde{\xi }^0 $
is not necessarily small but is \it smooth \rm (on length scale
$a_0$). The unknown error $e^0$
satisfies the equation
\be D_0 e^0 = r^0 . \label{Dee} \ee
%It involves the \it residual \rm
with  the residual $r^0 = f^0 -D_0\tilde{\xi }^0 $.
Given that $e^0$ is smooth, it can be obtained by smooth interpolation
of a suitable function $e^{1  } $ on $\Lambda^1$,
\be e^0 = \A^1 e^{1 }\ . \label {intp0}  \ee
That is, $e^0_z = \sum_{x \in \L^1} \A^1_{zx}e^{1  }_x $ with
$\A^1_{zx}$ which depends smoothly on $z$.
Now define a restriction operator $C^1$ such that
$C^1\A^1=1$.
Then~(\ref{intp0}) can be inverted,
 $ e^{1  } = C^1 e^0 \ . $
Applying $C^1 $ to both sides of~(\ref{Dee}) yields an equation for
$e^{1  }$,
\be D_1 e^{1  } = r^1 \ , \label{cgrideq}\ee
with
$ r^{1  } =  C^1r^0 $
and the effective operator $ D_1 = C^1D_0\A^1 $.
Given $e^{1  }$, one obtains $e^0$ from~(\ref{intp0}), and
$\tilde{\xi }^0 + e^0$ is an
improved solution of~(\ref{De}).
Thus, \it the problem has been reduced to an equation on the lattice
$\L^1 $ which has fewer points. \rm
If necessary, one repeats the procedure, moving to $\L^2 $ etc. The
procedure stops, because an equation on a ``lattice'' $\L^N$ with only a
single point is easy to solve.

The iterated interpolation
 $ \A^{[0j]} \equiv \A^1 \A^2 \dots \A^j$
                           from $\L^j$ to $\L^0$ should yield functions
on $\L^0 $ which are \it
           smooth on length scale $a_j$\rm , i.e. which change
little over a distance $a_j$ (in the ordered case).
For reasons of practicality, one must require
$ \A^j_{zx}= 0$ unless $z$ is near $ x$.
%-----------------------------------------------------------------------
%                      Section 3
%-----------------------------------------------------------------------
%
\section{SMOOTHNESS AND DISORDER}
 A successful MG scheme, whether
 deterministic or stochastic, needs smooth interpolation kernels $\A $.
 Thus we may ask: \it Which functions are smooth in
 the disordered situation, for instance in an external gauge field? \rm

 A (gauge covariant) naive answer is
\ba
 \sum_{\mu } (\nabla_{\mu }\xi , \nabla_{\mu }\xi )
 = (\xi , -\Delta \xi ) \geq \epsilon_0 (\xi , \xi )
\ea
 (with discretized covariant derivatives $\nabla_{\mu }$).
 By definition,
 the lowest eigenvalue $\epsilon_0$ of the negative
 covariant Laplacian $-\Delta $ is not small for disordered gauge
 fields. (It is positive and vanishes only for pure gauges.)
 Therefore there are no smooth functions in this case.

 Nevertheless there is an answer to the question, assuming
 a fundamental differential operator $D_0$ is specified by the problem
 (in the stochastic case, the Hamiltonian often provides
 $D_0$):

  \it A function $\xi $ on $\Lambda^0$ is smooth on length
 scale $a$ when \rm
 $\| D_0\xi \|^2 \ll \|\xi \|^2$ in units $a=1$.

 We  found that  a deterministic multigrid which employs
 interpolation kernels $\A^{[0j]}$ from $\L^j $ to the fundamental
lattice
 $\L^0$ which are smooth in this sense, works for arbitrarily disordered
 gauge fields.
 When there are no smooth functions in this sense at length scale $a_0$,
 then $D_0$ has no low eigenvalue, and there is no CSD
 and no need for MG.

 The above answer appears natural, and the
``projective MG'' of~\cite{10,11} is in its spirit.
 But to obtain kernels $\A^{[0j]}_{zx}$
 which are smooth \it on length scale $a_j$,  \rm
 one needs  approximate solutions of eigenvalue equations
 \be D_0 \A^{[0j]}_{zx} = \epsilon_0(x) \A^{[0j]}_{zx} \label{EVA}
 \ee  \rm
 Since $\A^{[0j]} $ is required to
 vanish for $z$ outside a neighbourhood of $x$, the  problem
 involves Dirichlet boundary conditions.
 For large $j$, $\A^{[0j]} $ will have a large support.
 If there is no degeneracy
 in the lowest eigenvalue, one can use inverse iteration combined
 with standard relaxation algorithms for the resulting inhomogeneous
 equation.
 But this and other standard methods will
 suffer from CSD again.
 Moreover, in the standard multigrid setup, one uses basic interpolation
 kernels $\A^j$ which interpolate from one grid $\L^{j}$ to the next
 finer one. In this case
  \be \A^{[0j]} = \A^1 \A^2\dots \A^j \ , \label{FACTOR} \ee
 and~(\ref{EVA}) becomes a very complicated set of nonlinear conditions.
 Possible solutions are
 \begin{description}
 \item[{\rm (i)}] Replace~(\ref{EVA}) by minimality of a cost functional
        (cp. later). Use neural algorithms to find kernels $\A^j $
        which minimize it.  This is still under study.
 \item[{\rm (ii)}] Give up factorization~(\ref{FACTOR}) and determine
        independent
        kernels $\A^{[0j]}$ as solutions of~(\ref{EVA}) by
        multigrid iteration. {\it This is done
        successively for $j=1,2,\dots$ One uses
        already determined kernels $\A^{[0k]}$ with $k<j$ for updating
        $\A^{[0j]}$.\/} We found that this works very well -
        cp. sect.~\ref{PERFORMANCE}.
 \end{description}
 Method (ii) will need of order $L^d \ln L$ storage space and
 $L^d\ln^2 L$ computational work for a $d$-dimensional system
 of linear extension $L$.
 %Method (ii) is not quite as efficient as
 %standard multigrid methods (MG) for ordered systems because of a large
 %overhead for storing and computing the kernels.
%-----------------------------------------------------------------------
%                      Section 4
%-----------------------------------------------------------------------
%
\section{CRITERIA FOR OPTIMALITY\label{CRITERIA}}
 Any iteration to solve~(\ref{De}) amounts to updating steps of the form
 \be \tilde{\xi}^0 \mapsto \tilde{\xi}^{ 0\prime } = \rho\
 \tilde{\xi}^0 + \sigma\ f^0 \ .
  \label{iterlin}
 \ee
 with the iteration matrix $\rho$ whose norm governs the convergence,
 and $\sigma= (1-\rho) D_0^{-1}$.
 If $\|\rho \|< 1$, the iteration converges with
 a relaxation time $ \tau \leq -1/\ln \|\rho \|$.
 \it Parameters \rm in the algorithm - such as operators
 $\A^j_{zx}$, $C^j_{xz}$ and $D_j$ -
 are \it optimal \rm if the cost functional
$ E = \|\rho\|^2 $ is at its minimum.

 As an example, consider a twogrid iteration in which a standard
 relaxation sweep on $\L^0$ with iteration matrix $\rho_0$
 is followed by exact solution of the
 coarse grid equation~(\ref{cgrideq}). The second step leads to
 an updating with some iteration matrix $\rho_1$, and
 $\rho = \rho_1\rho_0$. Therefore one may estimate
 $E\leq \|D_0\ \rho_0\|^2\ E_1 $ with
 $E_1 = \|D_0^{-1}\ \rho_1\|^2 $
 (fine grid relaxation smoothens the error
 but does not converge fast - therefore $\|D_0\ \rho_0\|$ is
 suppressed whereas $\|\rho_0\|$ is not much smaller than $1$)
 and try to  optimize the parameters
 above by minimizing $E_1$:
 Using the trace norm, $\|\rho \|^2 = {\rm tr}\ \rho \rho^{\ast }$,
 one finds
\ba
  E_1  =  {\rm Volume}^{-1}
          \sum_{z,w\in \L^0} |\Gamma_{zw}|^2
\ea
 with $\Gamma  =  D_0^{-1} - \A^1\ D_1^{-1}\ C^1$.
 Prescribing $C^1 $, and determining $D_1$ and $\A^1 $ by minimizing
 $E_1$ yields what we call the ``ideal interpolation kernel'' $\A^1_{zx}$
 for a given restriction map $C^1$. Since it has exponential tails
 instead of vanishing for $z$ outside a neighbourhood of $x$, it is
 impractical for production runs, though~\cite{12}.
%-----------------------------------------------------------------------
%                      Section 5
%-----------------------------------------------------------------------
%
\section{NEURAL MULTIGRID (NMG)}
 ~ A feed-for\-ward ar\-ti\-fi\-cial neu\-ral net\-work
 (ANN)~\cite{13} can perform the computations
 to solve~(\ref{Dee}) by MG relaxation.

\begin{figure}[htb] % h = here, tb = top or bottom p = page
 \epsfysize=170pt     % new y size
 \epsfbox[85 265 498 598]{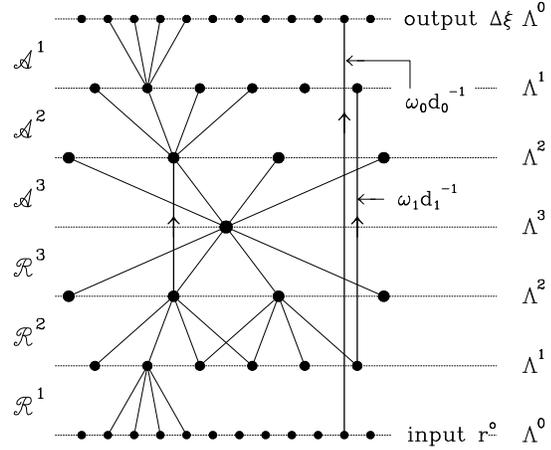}
 \caption{A feed-forward NMG architecture.}
 \label{FIGUR1}
 \end{figure}

 The nodes (``neurons'') of the NMG
 are identified with points of the MG as shown in Fig.~\ref{FIGUR1}.
 The resulting NMG consists of two copies of the same MG,
 except that the last layer $\L^N $ is not duplicated.
 In the standard MG approach, the basic interpolation kernels
 $\A^j $ interpolate from one layer $\L^j $ to the
 preceding one, $\L^{j-1}$.
 Each node is connected to some of the nodes in the preceding layer.
 In the upper half, the connection strength
 from $x\in \L^j $ to $z\in \L^{j-1}$ is $\A^j_{zx}$. In the lower half,
 node $z \in \L^{j-1} $ is connected to $x\in \L^j $ with strength
 $R^j_{xz}$. In addition there is a connection of strength
 $\omega_j d_j^{-1}$
 between the two nodes which represent the same
 point $z$ in $\L^j$ ($j<N$).

 According to Hebb's hypothesis of synaptical learning,
 a biological neural network learns by adjusting the
 strength of its synaptical connections.

 The network receives as input an approximate solution
 $\xi $ of~(2.1), from which the residual $r^0=f^0-D_0\xi $
 is then determined.
 It computes as output an improved
 solution $O = \xi + \delta \xi $. The desired output (``target'') is
 $\zeta = D_0^{-1}f^0$. $\delta \xi $ is a linear function of $r^0$.
 Except on the bottom layer,
 each node receives as input a weighted sum of the output of those
 nodes below it in the diagram to which it is connected. The weights
 are given by the connection strengths. Our neurons are linear because
 our problem is linear. The output of each neuron
 is a linear function of the input.

 The result of the computation is
 \be \delta\xi = ( \omega_0\ d_0^{-1}  +
 \sum_{k\geq 1} \A^{[0k]}\ \omega_k\  d_k^{-1} \ R^{[k0]})\ r^0 \ee
 where $R^{[k0]}=R^k...R^2R^1$ and $\A^{[0k]}= \A^1\A^2...\A^k$.
 The operators $C^j$ and $D_j (j>0)$,
 and the damping parameters $\omega_j$ $(j>0)$ are not needed
 separately since they only enter in the combination
 $R^j = C^j(1-\omega_{j-1}D_{j-1}d_{j-1}^{-1})$.
 The fundamental differential operator $D_0 $ and its
 diagonal part $d_0$ are furnished as part of the problem.
 The connection strengths (``synaptical strengths'') $\A^j_{zx}$,
 $R^j_{xz}$ (and possibly $\omega_0$) need to be found by a learning
 process in such a way that the actual output is as close as possible
 to the desired output.
 In supervised learning~\cite{13},
 pairs  $(\xi^{\mu }, \zeta^{\mu })$ (``training patterns'')
 are presented to an ANN.
 Given input $\xi^{\mu }$,
 the actual output $O^{\mu }$ is compared to the target $\zeta^{\mu }$,
 and the connection strengths are adjusted in such a way that the
 cost functional
 $E = \sum_{\mu } \|O^{\mu } - \zeta^{\mu } \|^2 $
 gets minimized. An iterative procedure to
 achieve this minimization is called \it  learning rule. \rm

 Taking for the sequence $\xi^{\mu }$ a complete orthonormal system
 of functions on $\L^0$, in the limit $f^0 \mapsto 0$,
 the target is $\zeta^{\mu }=0$ for any input, and the output
 $O^{\mu }= \rho\  \xi^{\mu }$ by~(\ref{iterlin}). The learning
 rule for the resulting cost functional
\ba E= \sum_{\mu } \|\rho\ \xi^{\mu }\|^2= {\rm tr}\ \rho \rho_{\ast }
  \equiv \|\rho \|^2 \stackrel{!}{=} \mbox{min}\
\ea
 is our previous optimality condition for multigrid relaxation
 in sect.~\ref{CRITERIA}.
%-----------------------------------------------------------------------
%                      Section 6
%-----------------------------------------------------------------------
%
\section{LEARNING RULE PERFORMANCE\label{PERFORMANCE}}
 The variant (ii) in sect.~\ref{SMOOTHNESS}
 involves a slightly different NMG. Instead of the
 connections between neighbouring layers of the multigrid,
 we now have
 connections from $\L^0$ to $\L^k$ with strength
 $C^{[k0]}_{xz}$, and from $\L^j$ to $\L^0$ with strength
 $\A^{[0j]}_{zx}$.
 If we adopt variational coarsening,
 all connection strengths are determined by
 interpolation kernels $\A^{[0k]}$ which have to be learned.
 The  damping factors $\omega_k$ were
 set to $1$ and $d_k$ is the diagonal part of $D_k$ as before,
 with
\be  D_k =
   \A^{[0k]\ast}D_0 \A^{[0k]}\ .
\ee
 The learning rule (ii) requires a process of ``hard thinking''
 by the NMG. Nodes which have learned their lesson
 already - i.e. which have their
 connection strengths fixed -
 are used to instruct the rest of the neural net, adjusting the
 strengths of the next layer of nodes in the NMG.

 A variant of this algorithm was tested in 2 dimensions, using
 SU(2)-gauge fields which were
 equilibrated with standard Wilson
 action at various values of $\beta$, and
 $D_0=-\Delta-\epsilon_0+\delta m^2 $.
 $\epsilon_0$ is the lowest eigenvalue of the covariant
 Laplacian $-\Delta $, and $\delta m^2 > 0$.
 Conventional relaxation algorithms for solving~(\ref{De}) suffer
 from CSD for such $D_0$, for any volume and
 small $\delta m^2$.

\begin{figure}[htb]
 \epsfxsize=215pt     % new x size
 \epsfbox[68 94 558 408]{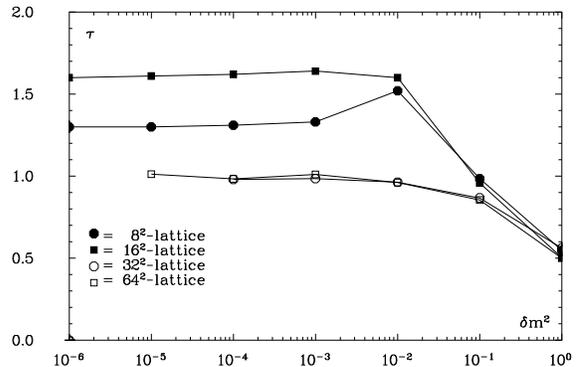}
 \caption{Correlation time $\tau$ as function of the lowest eigenvalue
 $\delta m^2$ in a representative gauge field configuration
 equilibrated at $\beta = 1.0$. For the $64^2$ lattice, $\tau$
 fluctuates very little with the gauge field.}
 \label{FIGUR2}
 \end{figure}

 It turned out that it was not necessary to find accurate solutions
 of the eigenvalue equation for the interpolation kernels $\A^{[0j]}$.
 An approximation $\A^{[0j]}_{zx}$ to  $(-\Delta )^{-n} \delta_{zx}$ was
 computed by multigrid iteration. It does not depend on $\delta m^2$.
 Updating  $\xi $ at $x \in  \L^j$ changes $\xi $ by
\ba
 \delta \xi_z=  \A^{[0j]}_{zx}d_{j,x}^{-1}r^j_x\ ,  \ \
 r^j=\A^{[0j]\ast }r^0 .
\ea
 The convergence rate (in units of MG iterations)
 of the $\xi$-iteration is
 shown in Fig.~\ref{FIGUR2} for $\beta =1.0$.
 One MG iteration involved one
 sweep (in checkerboard fashion) through each MG layer, starting with $j=0$.
%-----------------------------------------------------------------------
%                      Acknowledgements
%-----------------------------------------------------------------------
%
%\subsubsection*{ACKNOWLEDGEMENTS}

 We thank the HLRZ J\" ulich for computer time and assistance.

 Support by Deutsche Forschungsgemeinschaft is gratefully
 acknowledged.
%-----------------------------------------------------------------------
%                      References
%-----------------------------------------------------------------------
%

%
\end{document}